\documentclass[aps,prb,twocolumn,groupedaddress,showpacs,letterpaper,floatfix,10pt]{revtex4-1}

\usepackage{amsfonts,amsmath,amssymb}
\usepackage{graphicx}
\usepackage{natbib}
\usepackage[colorlinks=true]{hyperref}
\usepackage{newfloat}

\DeclareMathAlphabet{\mathitbf}{OML}{cmm}{b}{it}

\DeclareFloatingEnvironment[
    fileext=loa,
    listname=List of Algorithms,
    name=ALG.,
    placement=tb,
]{algorithm}

\def\ket#1{\vert #1 \rangle}
\def\bra#1{\langle #1 \vert}
\def\me#1#2#3{\bra{#1} #2 \ket{#3}}
\def\olap#1#2{\bra{#1} #2 \rangle}
\def\avg#1{\langle #1 \rangle}
\def\norm#1{\left\lvert #1 \right\rvert}
\def\mnorm#1{\left\lVert #1 \right\rVert}
\DeclareMathOperator{\Tr}{Tr}
\DeclareMathOperator{\Det}{Det}

\begin{document}

\title{Lazy skip lists, a new algorithm for fast hybridization-expansion quantum Monte Carlo}

\author{P. S\'emon}
\affiliation{D\'{e}partement de physique and Regroupement qu\'{e}b\'{e}cois sur les mat\'{e}riaux de pointe, Universit\'{e} de Sherbrooke, Sherbrooke, Qu\'{e}bec, Canada J1K 2R1}
\author{Chuck-Hou Yee}
\affiliation{Kavli Institute for Theoretical Physics, University of California Santa Barbara, CA 93106, USA}
\author{Kristjan Haule}
\affiliation{Department of Physics and Astronomy, Rutgers University, Piscataway, NJ 08854}
\author{A.-M. S. Tremblay}
\affiliation{D\'{e}partement de physique and Regroupement qu\'{e}b\'{e}cois sur les mat\'{e}riaux de pointe, Universit\'{e} de Sherbrooke, Sherbrooke, Qu\'{e}bec, Canada J1K 2R1}
\affiliation{Canadian Institute for Advanced Research, Toronto, Ontario, Canada, M5G 1Z8}

\begin{abstract}
  The solution of a generalized impurity model lies at the heart of electronic
  structure calculations with dynamical mean-field theory (DMFT). In the
  strongly-correlated regime, the method of choice for solving the impurity
  model is the hybridization expansion continuous time quantum Monte Carlo
  (CT-HYB). Enhancements to the CT-HYB algorithm are critical for bringing new
  physical regimes within reach of current computational power. Taking
  advantage of the fact that the bottleneck in the algorithm is a product of
  hundreds of matrices, we present optimizations based on the introduction and combination of two concepts of more general applicability: a) skip lists and b) fast rejection of proposed configurations based on matrix
 bounds. Considering two very different test cases with $d$ electrons, we find speedups of $\sim 25$ up to  $\sim 500$ compared to the direct evaluation of the matrix product. Even larger speedups are likely with $f$ electron systems and with clusters of correlated atoms.
\end{abstract}

\maketitle

\section{Introduction}
\label{sec:Intro}

One of the frontiers in condensed matter systems is the realistic modeling of
strongly-correlated materials. The combination of density functional theory
(DFT), a workhorse for electronic structure calculations of weakly-correlated
materials, with dynamical mean-field theory (DMFT) \cite{DMFTRMP}, originally designed to
handle strong correlations in simple models, has allowed insights into
strongly-correlated compounds at a level of realism previously
unobtainable. Comparisons of momentum-resolved spectral functions, densities of
states, and optics between theory and experiment are routine.

Lying at the core of this combined theory, named DFT+DMFT \cite{kotliarRMP, Aichhorn:2011, ChuckHaule, AmadonLDA+DMFT:2013, HeldReview2001, HeldReview:2007, LichtensteinLDA+DMFT}, is the solution of a
generalized Anderson impurity model. In the strongly-correlated regime, the
method of choice is the hybridization expansion continuous time quantum Monte
Carlo (CT-HYB)\cite{Werner:2006, Werner:2006General, hauleCTQMC, CTQMCRev}, a numerically exact algorithm capable of handling arbitrary
local interactions on the impurity site, in particular, the full atomic Coulomb
potential needed to capture the $d$ and $f$ electron physics present in
strongly-correlated materials. Enhancements to the CT-HYB algorithm are
important for bringing new physical regimes within the reach of current
computational resources.

In the context of model Hamiltonians, CT-HYB is also commonly used as an impurity solver for cluster generalizations of DMFT.~\cite{CDMFT:2001, QCT:2004, LTP:2006, HauleSC:2007, hauleAVOIDED, Park:2008, SentefSC:2011, SentefET:2011, Sordi:2010, Sordi:2011, Sordi:2012, Sordi:2012, Sordi:2013, ssht}  CT-HYB is particularly useful in the strongly correlated case.~\cite{gullCOMP}

\begin{figure}
\centering \includegraphics[width=\columnwidth]{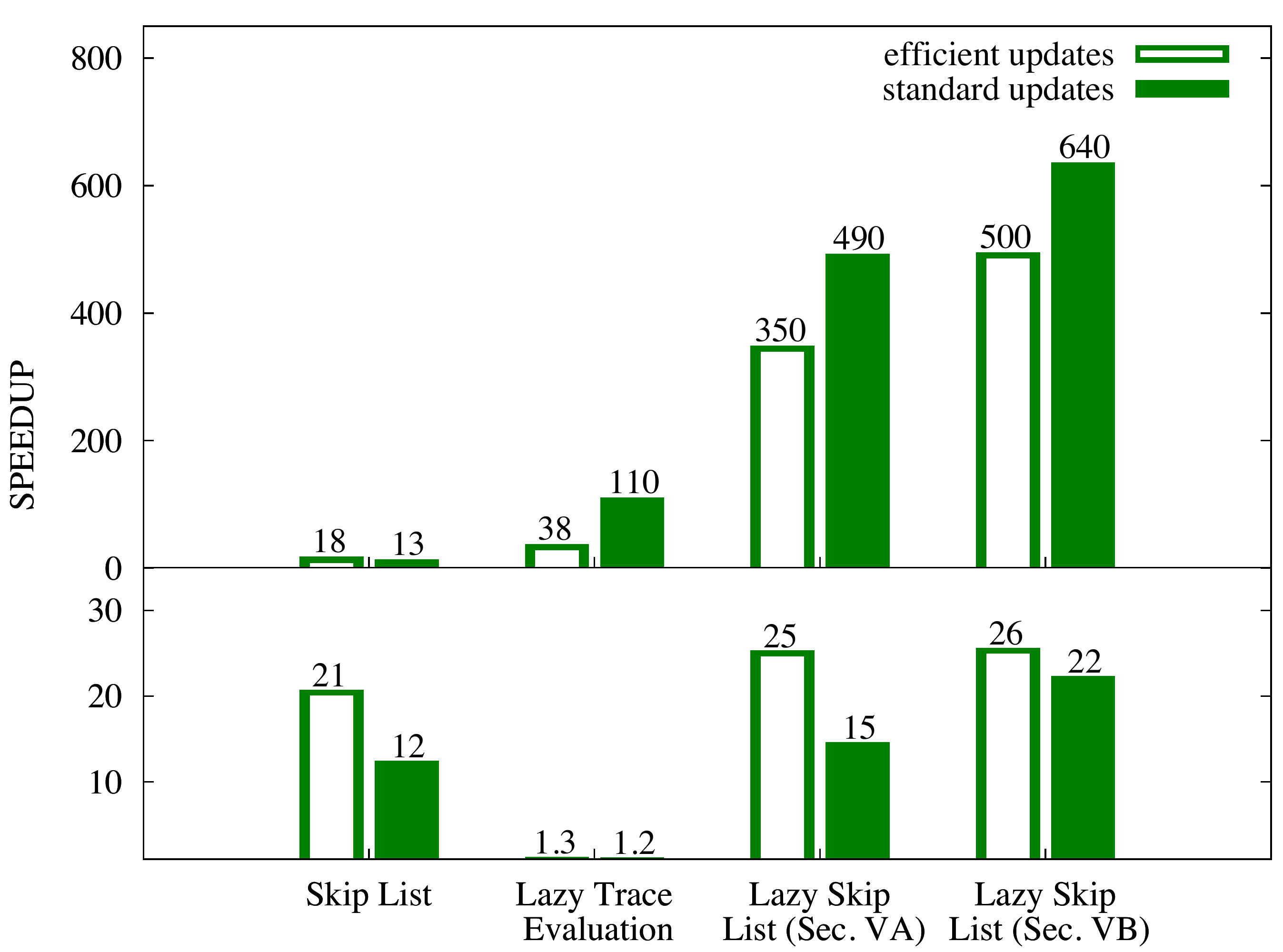}
\caption{Benchmark of different optimizations presented in this paper on the
  basis of a LNO thin film simulation \cite{Chuck} (top panel) and a FeTe simulation (lower panel), using standard updates with low acceptance ratio and efficient updates with high acceptance ratio. We measure the speedup of the skip lists (Sec.~\ref{sec:LazySkipListA} without lazy trace evaluation), the lazy trace evaluation (Sec.~\ref{sec:Lazy}) and the lazy skip lists (Sec.~\ref{sec:LazySkipListA} and Sec.~\ref{sec:LazySkipListB}), compared to a straightforward implementation (Sec.~\ref{sec:Symmetries}) as baseline.}
\label{fig:LNOFeSpeedup}
\end{figure}

Here, we present optimizations based on skip lists\cite{Pugh} and matrix bounds which
result in a speedup of $\sim 25$ up to $\sim 500$ as compared to the straightforward
implementation of CT-HYB (see Fig.~\ref{fig:LNOFeSpeedup}). These speedups are obtained for two very different test cases where the materials contain correlated $d$ electrons. In the low-temperature and strongly-correlated regimes
of interest, the most computationally expensive step is the evaluation of the
expectation value of a time-ordered sequence of (possibly thousands of) creation
and annihilation operators acting on the impurity degrees of freedom,
schematically notated as $\avg{d_1^\dagger d_2 d_3 d_4^\dagger d_5^\dagger d_6
  \cdots}$. When the complete basis of impurity states are inserted between
each operator, the problem is transformed into (the trace of) a product of
hundreds of matrices, called the impurity trace, which must be evaluated at
each Monte Carlo step.

Our algorithm, which we dub ``lazy skip lists'', optimizes the matrix product
by combining the following two ideas. First, we take advantage of the fact that between subsequent Monte
Carlo steps, the matrix product only changes by the insertion or removal of two
operators, for example, $\avg{d_1^\dagger d_2 d_3 d_4^\dagger d_5^\dagger d_6  \cdots} \rightarrow \avg{\mathitbf{d_i^\dagger} d_1^\dagger d_2 d_3 d_4^\dagger \mathitbf{d_j} 
  d_5^\dagger d_6 \cdots}$ in the case of insertion. We observe that the
intermediate products $d_1^\dagger d_2 d_3 d_4^\dagger$ and $d_5^\dagger d_6
\cdots$ are unchanged. Using skip lists, we efficiently store these
intermediate products to minimize recomputation. Historically, the expense of computing this matrix product led to optimizations, beginning with the left-right storage of intermediate products~\cite{hauleCTQMC}. This algorithm was of order $O(k)$, where $k$ is the order in perturbation theory. In Refs.~\onlinecite{GullThesis:2008,CTQMCRev} a faster binary-search-tree algorithm, scaling as $O(\log(k))$ was proposed. Skip lists are statistically as efficient as binary trees,~\cite{Pugh} better match the structure of the impurity trace and are simpler to implement.

Second, we often can avoid performing the matrix product altogether by quickly
rejecting proposed Monte Carlo moves via a ``lazy'' evaluation of the impurity
trace. This implementation was first carried out in Ref.~\onlinecite{yee-thesis} and already successfully used in Ref.~\onlinecite{AlphaPlutonium:2013}. In normal Monte Carlo sampling, we compute an acceptance probability $p$
for a proposed move, then accept the move if $p > u$, where $u$ is a number
chosen randomly in $[0,1]$. Here, we do the opposite: we flip the metaphorical
Monte Carlo coin to obtain $u$ first, then lazily refine bounds $p_\text{min} <
p < p_\text{max}$ on the acceptance ratio until $u$ drops outside the bracketed
interval. The bounding is fast, involving only scalar operations, and rapidly
converges because the time-evolution operators in the time-ordered operator
sequence often involve exponents which vary tremendously in magnitude.


We begin by reviewing the CT-HYB algorithm in Sec.~\ref{sec:CTQMC}, focusing on
the aspects relevant to this work. In the next two sections
(Sec.~\ref{sec:SkipList} and \ref{sec:Lazy}), we present independently the key
algorithmic advancements, skip lists and lazy trace evalution, which are
combined to form the final method in Sec.~\ref{sec:LazySkipList}. We benchmark
our optimizations in Sec.~\ref{sec:Results}. The Appendix explains how the trace can be bounded using matrix norms.  

\section{Continuous Time Quantum Monte Carlo}
\label{sec:CTQMC}

In this section, we briefly summarize the key steps which generate the
hybridization expansion formulation of impurity models. The goal is to quickly
arrive at a description of the structure of the impurity trace imposed by the physics and to discuss what it implies for the Monte Carlo algorithm.

A general impurity model consists of a local interacting system $H_\text{loc}$
describing the impurity degrees of freedom, immersed in a non-interacting
electronic bath:
\begin{multline}
  \label{equ:imp}
  H = H_{\text{loc}}(d_i^\dagger, d_i) + \sum_{\mu} \epsilon_{\mu} a_{\mu}^\dagger a_{\mu} \\
  + \sum_{i\mu}(V_{\mu i} a_{\mu}^\dagger d_i + \text{h.c.}),
\end{multline}
where $\epsilon_{\mu}$ is the bath dispersion and $V_{\mu i}$ the
amplitude for particles to hop from the impurity orbital $i$ to the bath
orbital $\mu$. The spin index is absorbed into the index $i$.

\subsection{Partition Function Sampling}
\label{sec:PartitionSampling}

In CT-HYB, we transform the partition function $Z = \Tr e^{-\beta H}$ of
the impurity model into a form amenable for Monte Carlo sampling (described in
detail in Ref.~\onlinecite{CTQMCRev}). One uses the interaction representation with the unperturbed Hamiltonian the sum of the local and bath Hamiltonians. The hybridization is the interaction term. Then, we expand the
resulting expression in powers of this hybridization term, giving
\begin{equation}
\begin{split}
  \label{equ:partitionsum}
  Z &= Z_{\text{bath}} \sum_{k=0}^\infty 
  \int_0^\beta \! d\tau_1  \cdots \int_{\tau_{k-1}}^\beta  \!\!\! d\tau_k
  \int_0^\beta \! d\tau_1' \cdots \int_{\tau_{k-1}'}^\beta \!\!\! d\tau_k' \\
  & \times \sum_{i_1\cdots i_k} \sum_{i_1'\cdots i_k'}
  w\{(i_1,\tau_1) \cdots (i'_k,\tau'_k)\},
\end{split}
\end{equation}
where the integrand is
\begin{multline}
  \label{equ:weights}
  w\{(i_1,\tau_1) \cdots (i'_k,\tau'_k)\} = \Det \boldsymbol{\Delta} \\
  \times \Tr_\text{loc} [\text{T}_\tau e^{-\beta H_{\text{loc}}}
    d_{i_k}(\tau_k) d_{i_k'}^\dagger(\tau_k') \cdots
    d_{i_1}(\tau_1) d_{i_1'}^\dagger(\tau_1')].
\end{multline}
Since the impurity and bath degrees of freedom are decoupled, the trace over the bath has been performed. The bath is contained
in the determinant of a $k\times k$ matrix $\boldsymbol{\Delta}$ with elements
evaluated from the hybridization function $(\boldsymbol{\Delta})_{mn} =
\Delta_{i_m'i_n}(\tau_m' - \tau_n)$ whose Matsubara definition is
\begin{equation}
  \Delta_{ij}(i\omega_n) =
  \sum_{\mu} \frac{V^*_{\mu i} V_{\mu j}}{i\omega_n - \epsilon_{\mu}}.
\end{equation}
The average over the impurity
$\Tr_\text{loc}$ in general cannot be further decomposed. Its evaluation
requires converting the sequence of operators (and intervening time-evolution
operators) into matrices in the basis of the impurity Hilbert space
$\mathcal{H}$.

The Monte Carlo sampling of Eq.~\ref{equ:partitionsum} proceeds as follows: the
integrands $w$ of the partition function sum define the weights of a
distribution over the configuration space $\lbrace (i_1, \tau_1) \dots (i_k',
\tau_k') \rbrace$ which is sampled with the Metropolis-Hastings algorithm. At
each step, a new configuration is proposed with probability $A$ and accepted
with probability
\begin{equation}
  \label{equ:prob}
  p = \min\left(1, \frac{A'|w|}{A|w'|}\right),
\end{equation}
where $w$ and $w'$ are the weights of the new and the old configuration respectively, and $A'$ is the proposal probability of the inverse update.

The bottleneck is that the weights $w$, and the expensive impurity trace
contained within, must be computed in order to decide whether to accept each
new proposed configuration. In terms of computational effort, if $N =
|\mathcal{H}|$ is the size of the local Hilbert space, and we are sitting at
perturbation order $k$, the impurity trace costs $O(N^3k)$ while the
hybridization determinant costs $O(k^3)$ (which can be reduced to $O(k^2)$ for local updates). The average expansion order $\avg{k}$, which is typically in the hundreds,
is proportional to the inverse temperature $\beta$, whereas the $N$ grows exponentially with the number of impurity
orbitals ($N=1024$ for the $d$-shell). Thus, except at very low temperatures, the calculation of the impurity trace is the bottleneck
in these Monte Carlo simulations.

Alluded to in the above discussion, the impurity trace contains a time-evolution
operator between each creation and annihilation operator, which we denote by
$P_\tau = e^{-\tau H_\text{loc}}$. We also write $(F_i)_{mn} =
\me{m}{d_i}{n}$ for the matrix representation of the creation and
annihilation operator, where $m$ and $n$ index the states in $\mathcal{H}$.  In
this notation, the impurity trace explicitly becomes an alternating matrix
product:
\begin{equation}
  \label{equ:operatorproduct}
  \Tr_\text{loc} P_{\beta - \tau_k} F_{i_k} P_{\tau_k - \tau_k'} F_{i_k'}^\dagger
    \cdots F_{i_1} P_{\tau_1 - \tau_1'} F_{i_1'}^{\dagger} P_{\tau_1'}.
\end{equation}
For simplicity, we have assumed that the imaginary times in
Eq.~\ref{equ:weights} are time-ordered as they appear.

\subsection{Symmetries, Sectors and Block Matrices}
\label{sec:Symmetries}

We can make a key simplification to the impurity trace using symmetries prior
to developing computational algorithms \cite{hauleCTQMC}. The local hamiltonian $H_\text{loc}$
generally possesses abelian symmetries (e.g. particle number, spin, momentum),
which allow us to decompose the impurity Hilbert space as a direct sum
$\mathcal{H} = \bigoplus_{q=1}^N \mathcal{H}(q)$. Here, $q$ enumerates the
sectors of the Hilbert space, each of which is characterized by a definite set of quantum numbers (e.g. particle number, spin, momentum).

Using these symmetries one defines a new basis for the creation-annihilation operators. A creation or annihilation operator, which we denote by a
generalized index $\alpha$ formed by combining its quantum numbers with
the type of operator (creation or annihilation), maps each sector $q$ either to
0 or uniquely to one other sector $q'$.  This leads to block
matrices $F_\alpha(q)$ which can be combined with a sector mapping function $s_\alpha$~\cite{hauleCTQMC} defined by
$s_\alpha(q) = q'$. The time-evolution operator maps each sector onto itself.

In the sector basis, the operator product in Eq.~\ref{equ:operatorproduct}
becomes $P F_{\alpha_{2k}} P F_{\alpha_{2k-1}} \cdots
F_{\alpha_2}PF_{\alpha_1}P$ that maps a sector $q_0$ onto $q_{2k}$ defined by the
string $q_0 \rightarrow q_1 := s_{\alpha_1}(q_0) \rightarrow \dots \rightarrow
q_{2k}:= s_{\alpha_{2k}}(q_{2k-1})$.  The impurity trace decomposes into a sum
over sector traces,
\begin{multline}
  \label{equ:sectortrace}
  \Tr P F_{\alpha_{2k}} \cdots F_{\alpha_1}P = \\
  \sum_{q_0} \Tr P(q_{2k}) F_{\alpha_{2k}}(q_{2k - 1})\cdots F_{\alpha_1}(q_0)P(q_0),
\end{multline}
and only sectors $q_0$ which are not mapped on $0$ contribute. Such mapping on $0$ generally occurs because of the Pauli principle. In a
typical 3$d$ impurity model with the full atomic Coulomb interaction, the
number of sectors is $\sim 100$ and the number of surviving strings ranges from
1 to $\sim 20$.

\section{Skip Lists}
\label{sec:SkipList}
We first begin with a motivation for skip lists. Then the skip list and the way it is used to store matrix sub-products is described. The final subsection explains how matrix multiplications can then be performed efficiently when operators are inserted or removed.  
\subsection{Motivation for Skip Lists}
At each Metropolis-Hastings step, a matrix product needs to be computed to decide 
whether the proposed configuration is accepted or rejected. One possibility is to always 
calculate all the products from scratch. However, only two matrices are typically 
inserted or removed, so this strategy is not only 
expensive, but also highly redundant. 

\begin{figure}[h]
\centering
\includegraphics[width=\columnwidth]{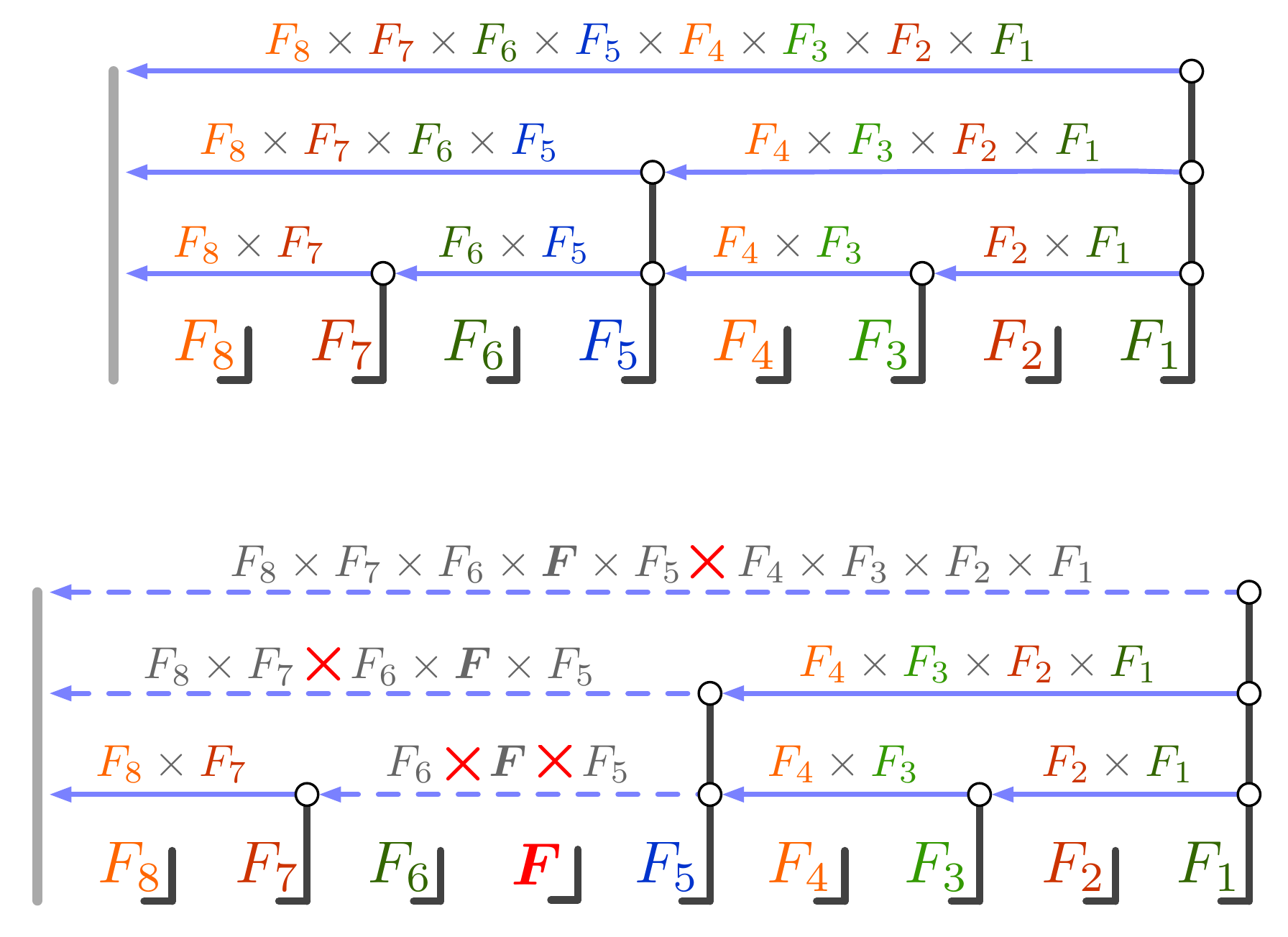}
\caption{Top panel: Storage scheme for sub-products of matrices. The arrows store 
the products of matrices they span over. The $l=1$ level stores the pair products, the 
$l=2$ their products and so on. Lower panel: The matrix $F$ has been inserted in the matrix 
product of the top panel and the products with a bold red multiplication sign need 
to be calculated in order to obtain the total product.}
\label{fig:BalancedList}
\end{figure}

To avoid multiplying almost all the time the same matrices, we may pair them off 
and store their product. This way almost every second 
multiplication is skipped when calculating the product of a proposed configuration. 
However, this is not yet optimal. One can store products of four, eight matrices etc. leading to a collection of sub-products that will allow us to minimize the number of redundant multiplications. 
This storage strategy may be represented as shown in Fig.~\ref{fig:BalancedList}, where we omit 
the propagators for simplicity.
The arrows store the sub-products of operators they span, including the operator 
they start from and excluding the operator they point to.

Inserting now a matrix $F$, some of the stored sub-products expire, as shown on the lower panel of 
Fig.~\ref{fig:BalancedList}. These are the sub-products of arrows that span over 
the inserted matrix. To calculate the product of the proposed configuration, we begin 
with the arrow just above the inserted operator. This costs two multiplications, $F_6\cdot F\cdot F_5$. 
Moving up, the next missing sub-product $F_8F_7\cdot F_6FF_5$ is calculated from the two 
sub-products below with one multiplication, and multiplying this sub-product with $F_4F_3F_2F_1$ 
yields the total product. Except at the first level, this involves one matrix multiplication per level, 
as each arrow is the product of two arrows one level below. For 32, 128 and 512 operators, a 
representation like that in Fig.~\ref{fig:BalancedList} has 5, 7 and 9 levels respectively, 
and the number of matrix multiplications is logarithmic in the number of operators in 
the product.  However, this storage scheme works only if the expansion order is a 
power of two, and we have to find a strategy to maintain an equilibrated structure 
when inserting or removing matrices at random places. Equilibrated means that a sub-product 
is ideally always the product of two sub-products one level below. 

For simplicity, we ignore here the block structure of the operator matrices. Their discussion is postponed to
Sec. ~\ref{sec:LazySkipList}.

\subsection{Skip Lists and Matrix Products}
\label{sec:SkipListA}
In Fig.~\ref{fig:BalancedList}, the heights of the vertical bars associated with the matrices 
organize the arrows, that is the sub-products. The original matrices are stored at level $l=0$. There is an arrow starting and ending 
at the top end of each bar with level $l>0$, except for the first bar on the right where no arrow ends. When inserting an 
operator, we are free to associate a bar to this operator at a height that we may choose.  The choice of skip lists\cite{Pugh} 
is to take a height $l$ that is determined randomly according to the distribution $2^{-l-1}$,
that is, half of the bars are
\begin{figure}[h]
\centering
\includegraphics[width=\columnwidth]{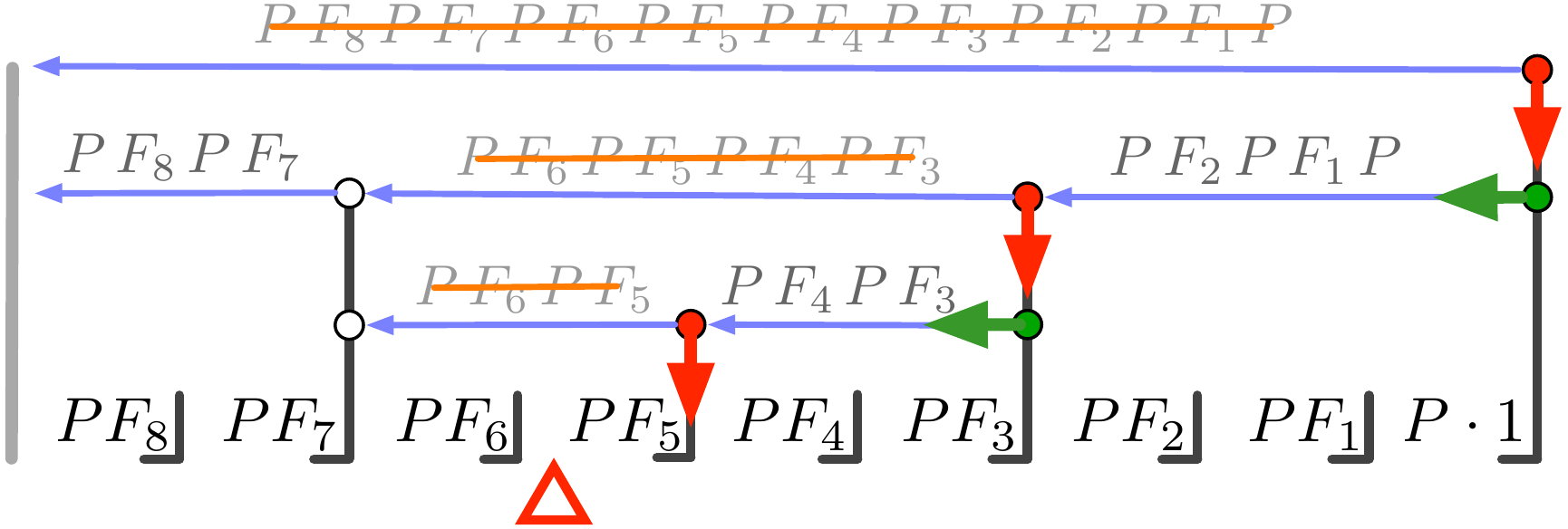}
\caption{Skip list to store sub-products of operators $F_i$ and propagators $P$. 
The arrows store the products they span over. The bold arrows in red and green show the path that is followed
when a matrix is inserted at the place indicated by the red triangle. The products stored in the blue arrows are emptied if their tail coincides with that of the bold red arrows.}
\label{fig:SkipList}
\end{figure}
on average at least level one, a quarter at least level two, and so on. This keeps 
the skip list on average equilibrated. A typical arrangement is shown in Fig.~\ref{fig:SkipList}. 
Here we include the propagators, and an arrow stores the 
sub-product starting with the operator at its tail and ending with the propagator 
at its head. However, to include the first propagator $P$ appearing on the right, we need to store the product of $P$ with the identity matrix at the first bar. Since the heights are chosen randomly, there is no guaranty that the height of that first bar 
exceeds all others as in Fig.~\ref{fig:BalancedList}. Hence we just assume that it is at a height that exceeds all others.  

To calculate the product after insertion of one operator in this skip list, we can proceed 
as in Fig.~\ref{fig:BalancedList} if the randomly chosen height of the associated bar is zero. This changes 
if the height is not zero. More importantly, two operators and sometimes more must be inserted or removed 
at once in Monte-Carlo simulations,~\cite{SemonErgodicity:2014} whereas the product is needed at the end only. Also, 
combinations of insertions and removals are sometimes necessary to make the sampling 
more efficient. Hence, we need a flexible multiplication algorithm, which is discussed in 
the next section. 

\subsection{Skip Lists and Matrix Multiplication}
\label{sec:SkipListB}
To calculate the new product after an arbitrary sequence of insertions and/or removals 
with a minimal number of matrix multiplications, we proceed in two steps. First the matrices
are inserted and/or removed, one after the other. At each time, this invalidates some sub-products $M=PF....PF$, stored in the blue arrows.  
These sub-products are thus emptied. Once the new configuration is proposed, the product is calculated 
by filling up the emptied sub-products.

When inserting an operator in the skip list, a sub-product expires if the operator lies 
between the head and the tail of the corresponding arrow, see Fig.~\ref{fig:SkipList}. To identify all 
such arrows, we follow the skip list insertion algorithm~\cite{Pugh} and begin at the 
tail of the top arrow. This arrow necessarily spans over the operator to insert, and its 
sub-product is emptied. Moving down the red arrow on the right in Fig.~\ref{fig:SkipList} to the next lower blue arrow, we test if the operator to 
insert lies between the head and tail of this arrow. If yes, the sub-product is emptied, 
and the next lower blue arrow is tested. If not, the arrow is traversed and the process is repeated until we end up by emptying the sub-product at the 
blue arrow just above the place where the operator will be inserted. Proceeding likewise for removal, all expired 
sub-products are emptied once the new configuration is proposed\footnote{Arrows starting 
from the bar of an inserted operator are always empty.}.

To fill up the emptied sub-products $M$ once the insertions and/or removals are completed, we proceed recursively. 
The sub-product at an  arrow $\mathcal{A}$ can be calculated from the sub-products 
$M_a,M_{a+1},\dots M_b$ stored at the arrows $\mathcal{A}_a,\mathcal{A}_{a+1},\dots,\mathcal{A}_{b}$ 
just below. If all of these sub-products have not been emptied, they are multiplied while traversing the 
arrows $\mathcal{A}_a\rightarrow \mathcal{A}_{a+1} \rightarrow \cdots$ and the result is stored at the 
arrow $\mathcal{A}$. If however one of the sub-products $M_i$ at an arrow $\mathcal{A}_i$ is missing, 
we recursively calculate this sub-product from the sub-products below the arrow $\mathcal{A}_i$. This recursion 
stops at the latest at the bottom of the skip list, where the operators are multiplied with the propagators. The total product
is obtained by starting the recursion at the top arrow.

Once the new product is calculated, we decide whether to accept or reject the proposed configuration. 
To recover the skip list in case of rejection, a backup is taken at the beginning of a trial step.

\section{Lazy Trace Evaluation}
\label{sec:Lazy}

In the regimes of interest (moderate to low temperatures $T\lesssim 100$~K,
strong Coulomb interaction $U \gtrsim 5$~eV), the probability of accepting a
proposed move is low, generally lying below 10\% and often below 1\%. The Pauli
principle and time-evolution operators $e^{-\Delta\tau H_\text{loc}}$ place
strong constraints on the insertion/deletion of operators, causing the low
acceptance probabilities. Developing techniques to reject improbable moves with
minimal computational effort is crucial.

The Pauli constraint is computationally neglegible, as it can quickly be
determined by following the string of sector mappings $q_0 \rightarrow q_1
\rightarrow q_2 \cdots$ and checking that not all strings are annihilated
(i.e. mapped to 0). In contrast, the time-evolution operators are interspersed
within the matrix product. Proposed moves often drive transitions to
high-energy sectors, where the exponentials $e^{-\Delta\tau H_\text{loc}}$
strongly suppress the acceptance probability. Here, we describe a ``lazy
trace'' algorithm which leverages these exponentials to efficiently reject
moves with low acceptance probability, largely avoiding a full evaluation of
the impurity trace.

\begin{figure}
  \includegraphics[width=\columnwidth]{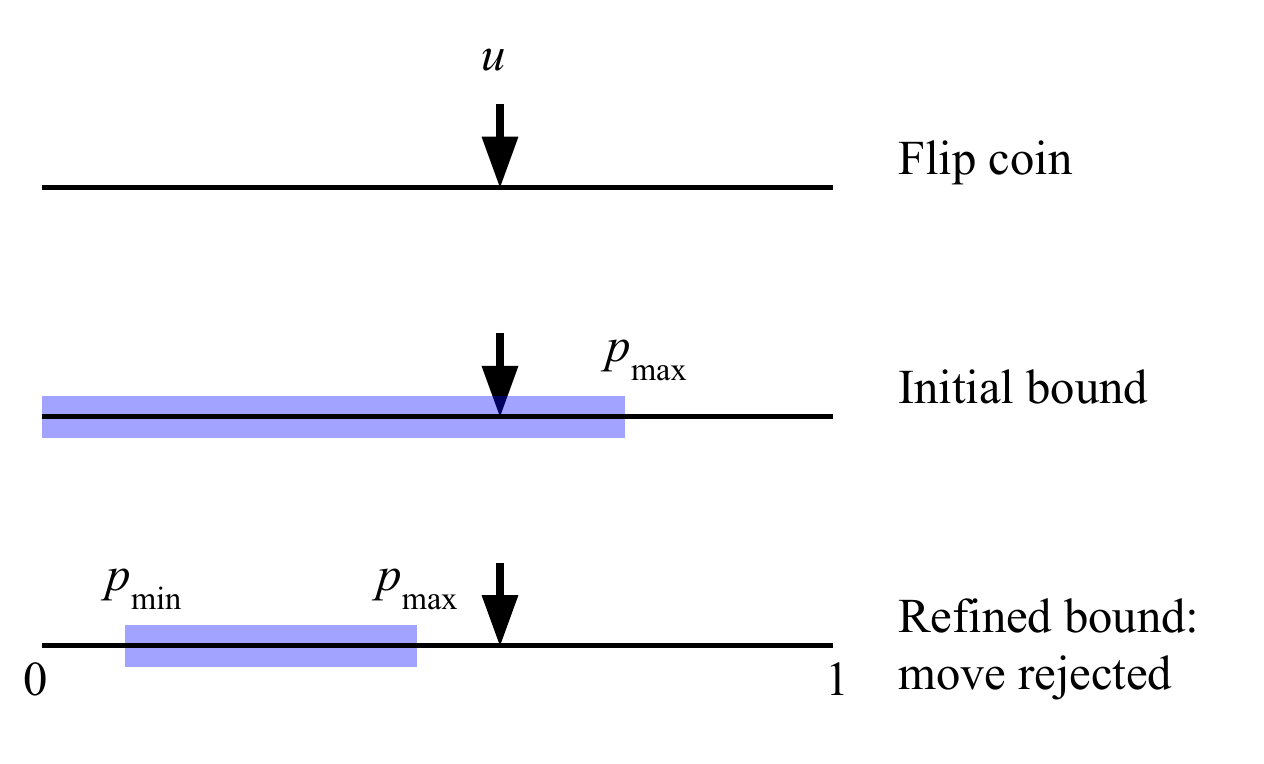}
  \caption{The bounding technique within the lazy trace evaluation. We first
    flip a coin to obtain a random number $u \in [0,1]$. Then, using
    sub-multiplicative matrix norms, we compute initial bounds $p_\text{min} <
    p < p_\text{max}$ on the acceptance probability. The bounds are refined
    until $u$ falls outside $[p_\text{min}, p_\text{max}]$ and the move can be
    definitively accepted or rejected.}
  \label{fig:bounding}
\end{figure}

The first component of the lazy trace algorithm~\cite{yee-thesis} is fast bounding of the
impurity trace in each symmetry sector. Writing in shorthand
Eq.~\ref{equ:sectortrace} as $\Tr = \sum_q \Tr_q$, \emph{assume} we can quickly
compute bounds $B_q \ge |\Tr_q|$ for each sector trace. This provides a maximum
bound on the trace via the triangle inequality:
\begin{equation}
  |\Tr| \le \sum_q |\Tr_q| \le \sum_q B_q.
\end{equation}
Using the expression for the acceptance probability $p$ (Eq.~\ref{equ:prob}),
and writing the weight of the old configuration as $w' = \Det' \cdot \Tr'$, we
obtain an upper bound
\begin{equation}
  p_\text{max} = \frac{A'}{A} \frac{|\Det|\sum_q B_q}{w'}. 
\end{equation}
This bound can be refined as follows: take the sector $q_\text{max}$ with the
largest $B_q$ and compute the exact sector trace $\Tr_{q_\text{max}}$. Applying
the reverse triangle inequality gives
\begin{equation}
  \Big| |\Tr| - |\Tr_{q_\text{max}}| \Big| \le \sum_{q \neq q_\text{max}} B_q,
\end{equation}
producing refined bounds
\begin{equation}
  \begin{pmatrix} p_\text{max} \\ p_\text{min} \end{pmatrix}
  = \frac{A'}{A} \frac{|\Det|}{w'} \left( \norm{\Tr_{q_\text{max}}} \pm \sum_{q\neq q_\text{max}} B_q \right).
\end{equation}
This procedure can be continued, generating successively tighter bounds, until
we obtain the exact trace. The sequence of bounds is likely to tighten most
rapidly if we choose the sectors in decreasing order of $B_q$.

The second key idea is to flip the Monte Carlo coin first to obtain the
acceptance threshold $u$, before computing the above approximation to the
acceptance probability. If $p_\text{max} < u$, and it often is, we can reject
the move outright. If $p_\text{min} > u$ we accept the move. If neither of these possibilities occur, we successively refine the bounds on $p$ until we
can either accept or reject the move, as illustrated in
Fig.~\ref{fig:bounding}.  In the following, we describe the construction of the
bounds $B_q$.

The basic equation is the formula
\begin{equation}
  \label{equ:norm}
  \left| \Tr A_1 A_2 \cdots A_n \right| \le C \cdot \mnorm{A_1} \mnorm{A_2} \cdots \mnorm{A_n},
\end{equation}
proven in Appendix~\ref{apdx:trace}. Here $A_k$ are matrices (not necessarily
square, although the entire product must be), $\mnorm{\cdot}$ is a
sub-multiplicative matrix norm, and $C$ is a constant which depends on the
specific matrix norm chosen and the dimension of the matrices. In the lazy trace algorithm, the
spectral norm (see
Appendix~\ref{apdx:trace}) is used. For rectangular matrices $A_l \in \mathbb{R}^{N_l\times M_l}$, the constant $C$ becomes the dimension of
the smallest matrix within the product, $C = \min\{N_l\}$. The spectral
norm is unity for a creation or annihilation operator, and $e^{-\Delta\tau_i
  E_0(q_i)}$ for time-evolution operator, where $E_0$ is the ground state
energy of the sector $q_i$ and $\Delta\tau_i$ is the time spent in this sector.

Application to the trace of a single sector in Eq.~\ref{equ:sectortrace} gives
\begin{multline}
  \label{equ:simplesectornorm}
  \norm{\Tr P(q_{2k}) F_{\alpha_{2k}}(q_{2k - 1})\cdots F_{\alpha_1}(q_0)P(q_0)} \\
  \le \min\{\dim\mathcal{H}(q_i)\} \cdot \exp \left(-\sum_{i=0}^{2k} \Delta\tau_i E_0(q_i) \right),
\end{multline}
While extremely cheap to calculate, this bound precisely captures the vast
variations in magnitude caused by exponentials in the time-evolution operators. The bounds for
each sector $B_q$ decrease extremely rapidly; in many cases, the
initial $p_\text{max}$ is sufficient to reject a proposed move.

When a move is accepted, the trace needs to be evaluated exactly, up to numerical accuracy, to be able to compute the acceptance probability of the next move. 

\section{Lazy Skip Lists}
\label{sec:LazySkipList}
In this section, we begin by combining the algorithms presented in Sec.~\ref{sec:SkipList} and Sec.~\ref{sec:Lazy}. In a second step, we show how 
the bounds on the sector traces in Sec.~\ref{sec:Lazy} may be improved using this combined algorithm. 
\subsection{Skip Lists and Lazy Trace Evaluation}
\label{sec:LazySkipListA}
When iteratively refining the bounds in the lazy trace evaluation, we only need the contribution to the trace
of one sector $q_0$ at a time in Eq.~\ref{equ:sectortrace}. To achieve this with the skip lists in Sec.~\ref{sec:SkipListA}, 
we begin by taking into account the block structure of the matrices. 

The operators $F$ and the sub-products $M$ are stored in their block form as pairs $s(q),F(q)$ and $s(q),M(q)$ of mapped 
sectors and corresponding matrix blocks. Similar to the total product which splits into strings in 
Sec.~\ref{sec:Symmetries}, this splits a sub-product $PF_b\cdots PF_a$ into sub-strings 
$P(q_{b+1})F_b(q_b)\cdots P(q_{a+1})F_a(q_a)$. Such a sub-string is stored in the matrix 
block $M(q_a)$ together with the mapped sector $s(q_a) := q_{b+1}$. 

To calculate one string in the total product, we only need one of the sub-strings of a given sub-product. When recursively
updating the sub-products in the skip list as in Sec.~\ref{sec:SkipListB}, we thus have to specify at each 
arrow $\mathcal{A}$ the requested sub-string by a start sector $q_a$. To select the entries in the block matrices $M_i$ (stored in $\mathcal{A}_i$ below $\mathcal{A}$) which need to be multiplied to obtain the requested sub-string $M_b(q_b)\cdots M_{a+1}(q_{a+1})M_{a}(q_{a})$, one maps the start sector $q_a$ into $q_{b}$ using the sector mappings $s_i$
at the arrows $\mathcal{A}_i$, namely 
$q_a \rightarrow q_{a+1} := s_a(q_a) \rightarrow \cdots \rightarrow q_b :=s_{b-1}(q_{b - 1})$.
The product is then stored in the matrix block $M(q_a)$ at the arrow $\mathcal{A}$, together 
with the mapped sector $s(q_a):=q_{b+1}$. Again, if a matrix block $M_i(q_i)$ at an arrow $\mathcal{A}_i$ is empty, 
we proceed recursively.

The combination of the skip lists and the lazy trace evaluation is now straightforward. First, expiring sub-strings are emptied when 
inserting and/or removing operators in the skip list, similar to Sec.~\ref{sec:SkipListB}.
Once the new configuration has been proposed, we start the recursion at the top arrow of the skip list separately for each sector 
needed by the lazy trace evaluation.

\subsection{Sub-products and Trace Bounds}
\label{sec:LazySkipListB}
The bounds on the sector traces in Eq.~\ref{equ:simplesectornorm} are calculated from the product of the norms of each
propagator and operator individually. Tighter bounds may be obtained by using the norms of stored sub-products. 
In Fig.~\ref{fig:BalancedList} for example, the trace is bounded by  
\begin{equation}
|\text{Tr}|  \le C \cdot \lVert F_8 F_7\rVert \lVert F_6 \rVert \lVert F \rVert \lVert F_5 \rVert \lVert F_4F_3F_2F_1 \rVert
\end{equation}
after insertion of the matrix $F$. Such bounds for a given sector trace $\text{Tr}_q$ are obtained recursively, in a manner analog to the block-matrix product 
of the corresponding string. 

Calculating the spectral norm of a stored matrix block is expensive, so the Frobenius norm 
is used here instead. While this norm is larger than the spectral norm, its numerical cost is small compared to a matrix 
multiplication. However, this means that this bound is not necessarily smaller than the one in Sec.~\ref{sec:Lazy}. Other choices for the norms are discussed in Appendix~\ref{apdx:trace}

\section{Two examples}
\label{sec:Results}

In this section we benchmark the skip lists (Sec.~\ref{sec:SkipList} taking into account the block structure described in Sec.~\ref{sec:LazySkipListA}), the lazy trace evaluation 
(Sec.~\ref{sec:Lazy}) and the lazy skip lists (Sec.~\ref{sec:LazySkipListA} and Sec.~\ref{sec:LazySkipListB}).
To this end, we consider Anderson impurity problems that appear in DFT+DMFT electronic structure calculation for thin film of LaNiO$_3$ (LNO)~\cite{Chuck,LNOprb:2011} and FeTe bulk compound~\cite{ChuckHaule}, using experimental structure of Ref.~\onlinecite{May:2010}  and Ref.~\onlinecite{FeTe-prb:2011}, respectively

In both cases, the impurity is a d-shell system, and the associated Hilbert Space splits into 132 sectors. The Slater parametrization of the Coulomb interaction is used. The average expansion orders 
are $\langle k \rangle \approx 225$ for LNO and $\langle k \rangle \approx 515$ for FeTe. The benchmarks are performed using 
two kinds of Metropolis-Hastings updates: i) standard ones,\footnote{Two operators are inserted anywhere between $0$ and 
$\beta$.} with low acceptance ratio and ii) efficient ones,\footnote{Two operators $d_i$ and $d_i^\dagger$ with given orbital and spin index $i$ are inserted between two consecutive 
operators with the same orbital and spin index $i$, not taking into account the position of operators with other orbital and spin indices. Both orderings $d_id_i^\dagger$ and 
$d_i^\dagger d_i$ of the inserted operators lead to a finite trace. These updates are in principle
ergodic and give the same results as the standard updates, however with less noise for fixed amount of CPU-time. The 
acceptance ratio is about 10-25 times higher.}   with acceptance ratio higher by a factor 10 to 25.

\begin{figure}
\centering \includegraphics[width=\columnwidth]{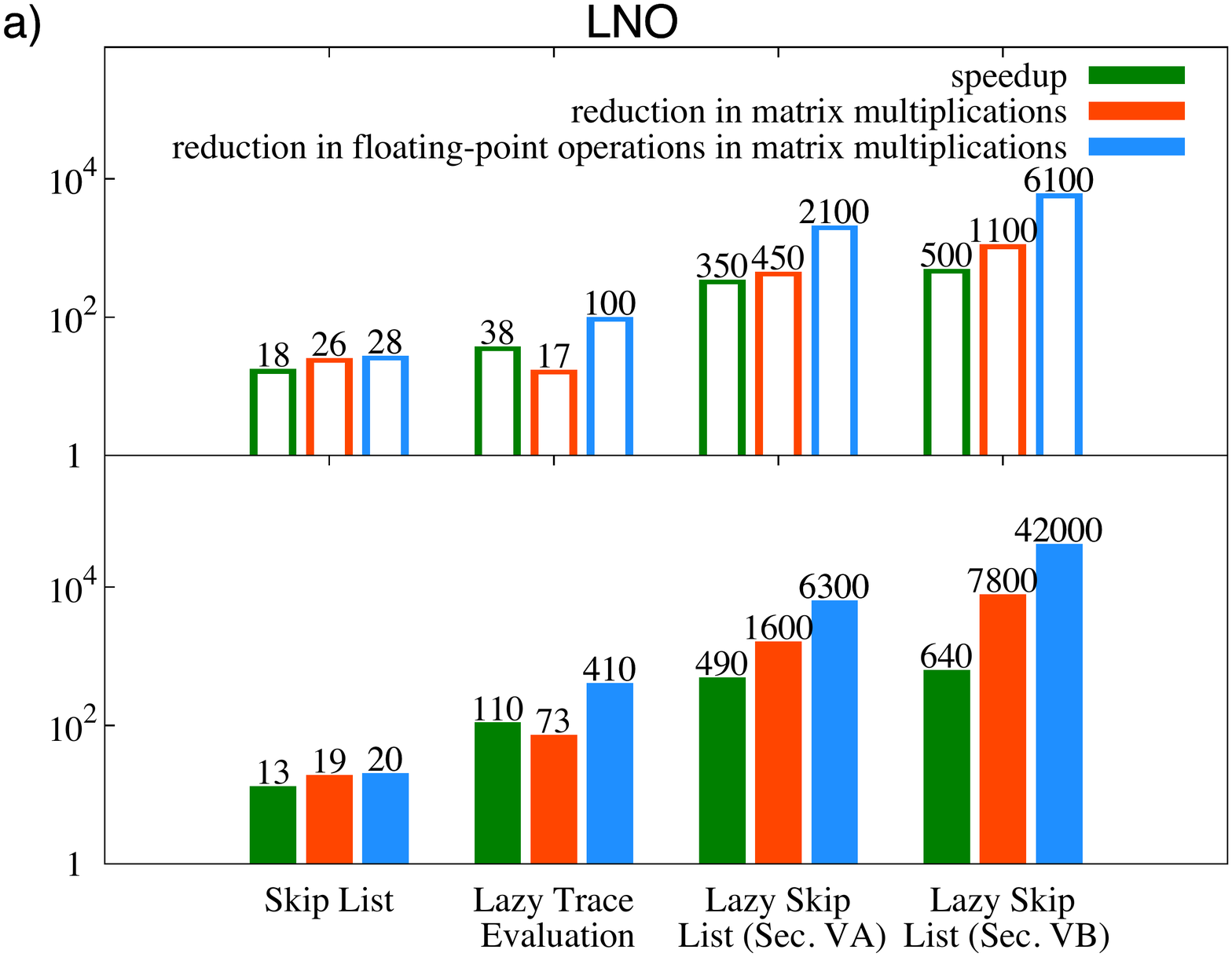}
\centering \includegraphics[width=\columnwidth]{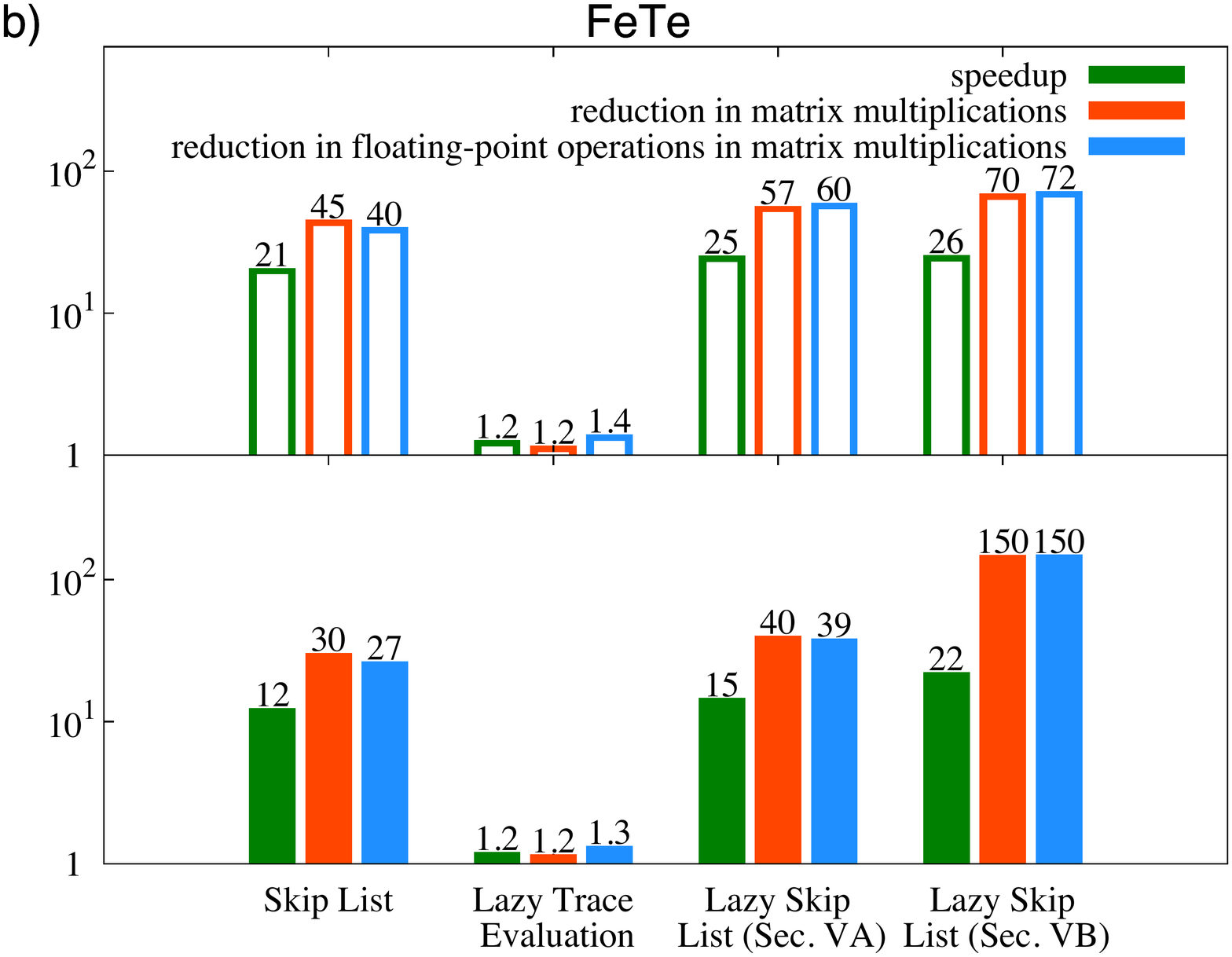}
\caption{Benchmark of different optimizations presented in this paper on the
  basis of a LNO thin film simulation (a) and a FeTe simulation (b): using efficient updates with high acceptance ratio (top panel) and 
  standard updates with low acceptance (lower panel). We measure speedup, reduction in
  matrix-multiplications and reduction in floating-point operations within matrix-multiplications, 
  with a straightforward implementation (Sec. \ref{sec:Symmetries}) as baseline.}
\label{fig:Combo}
\end{figure}

Fig.~\ref{fig:LNOFeSpeedup} shows the speedups of the different optimizations presented in this paper compared with, as a baseline, a straightforward
implementation (Sec.~\ref{sec:Symmetries}) that takes the block structure into account. Note the logarithmic scale. The skip lists alone accelerate the simulations for both test cases by a factor of about 20. 
While the lazy trace evaluation gives a substantial speedup for LNO, essentially no speedup is obtained for FeTe. This 
also shows in the performance of the combined algorithms, the lazy skip lists, which, with speedups of order 500, perform much better for LNO. The reasons for this difference between LNO and FeTe will become clear below. 

Fig.~\ref{fig:Combo} shows, in addition to the speedup, the reduction 
in matrix multiplications and the reduction in floating point operations. While combining 
different optimizations does not always result in an additional speedup, in our case 
the lazy trace evaluation and the skip lists work well together. The 
reduction in matrix multiplications for the lazy skip lists (Sec.~\ref{sec:LazySkipListA}) is essentially the product of the reductions for the 
lazy trace evaluation and the skip lists separately. While the reduction in matrix multiplications for the lazy 
skip lists in Sec.~\ref{sec:LazySkipListB} is less evident to anticipate, there is always an additional speedup 
that comes from calculating the bounds using the norms of the stored sub-products in the skip list. 

Note that speedups are smaller than expected from 
the reduction in matrix multiplications and floating point operations, in particular for the lazy skip lists 
of Sec.~\ref{sec:LazySkipListB}. This is due to the optimization overhead and to the fact that other parts than the local trace evaluation in the CT-HYB 
expansion, such as the evaluation of the determinants, are beginning to take a significant proportion of the total time. 

\begin{figure}
\centering \includegraphics[width=\columnwidth]{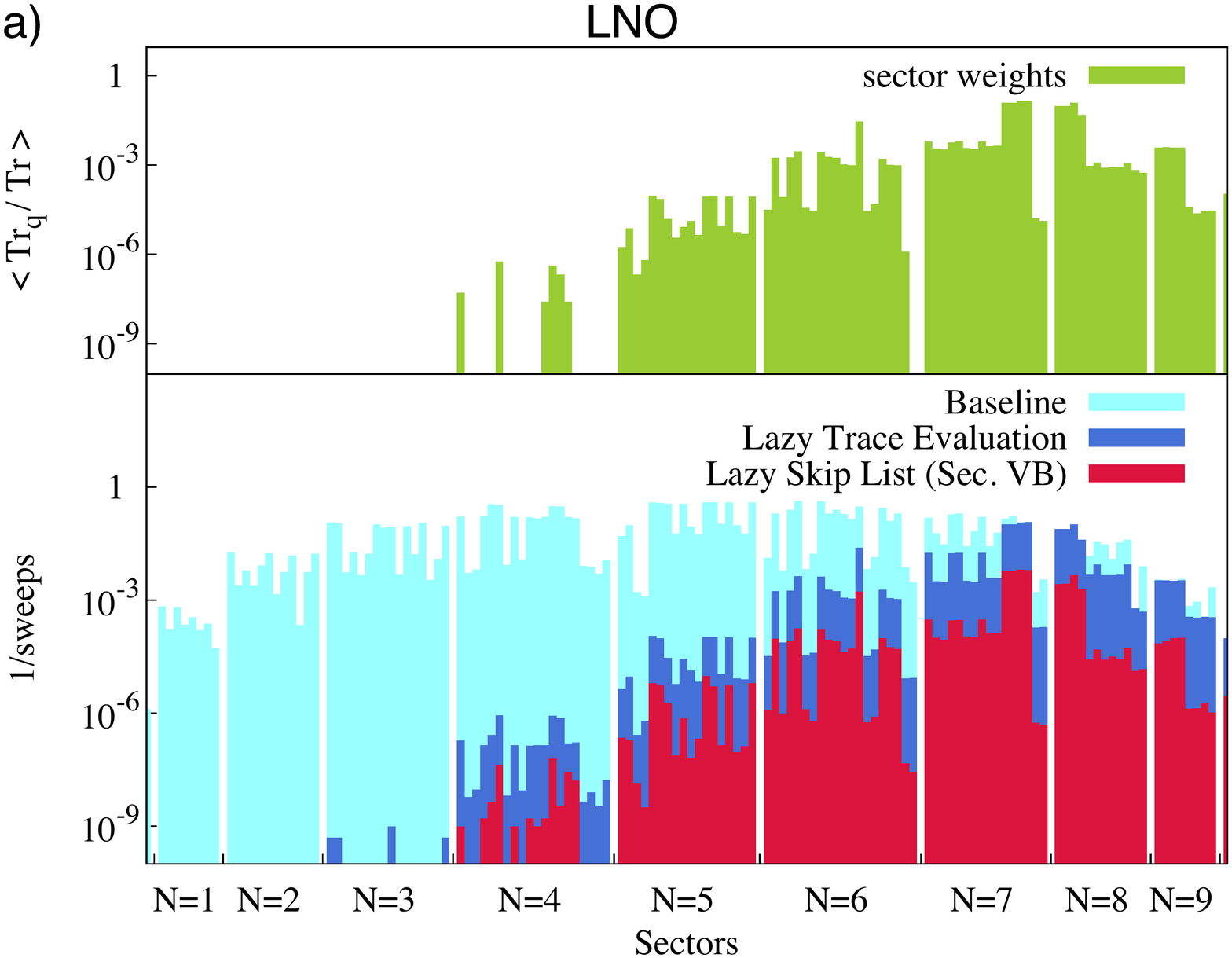}
\centering \includegraphics[width=\columnwidth]{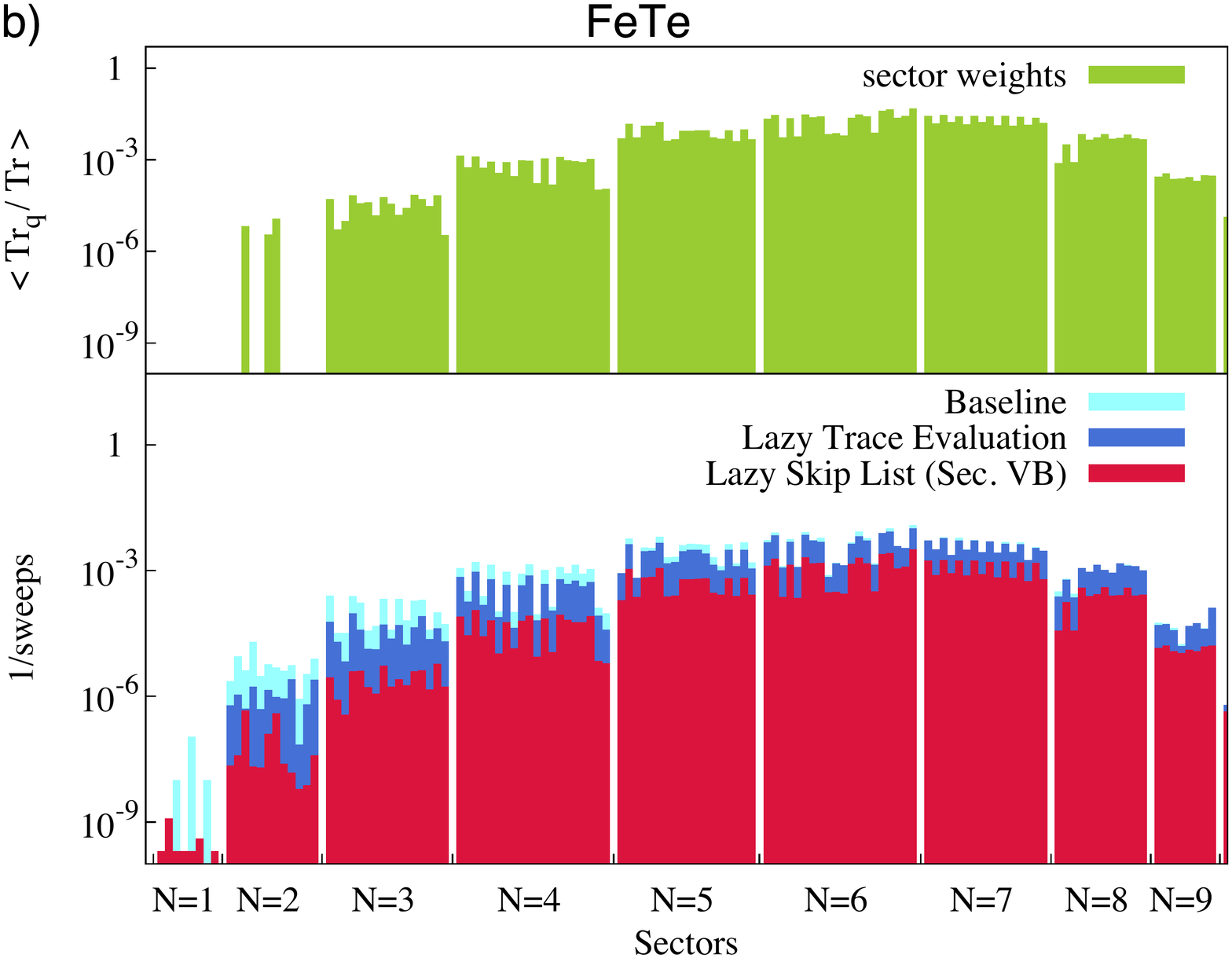}
\caption{On the basis of a LNO thin film simulation (a) and of a FeTe simulation (b) with standard updates: average weight $\langle \text{Tr}_q / \text{Tr} \rangle$ of a sector $q$ in the partition function expansion (top panel) and frequency with which $\text{Tr}_q$ is calculated 
for a sector (lower panel).}
\label{fig:Statistics}
\end{figure}

To understand why most of the speedup comes from the lazy trace evaluation for LNO while it comes from the skip list for FeTe, it is useful to consider the sector weights. We use standard updates. In Fig.~\ref{fig:Statistics}a) we show results for LNO and in Figs.~\ref{fig:Statistics}b) results for FeTe. Note the logarithmic vertical scales. The top panels display the average weights $\langle \text{Tr}_{q}/\text{Tr} \rangle $ 
of the various sectors in the partition function expansion. The lower panels of Figs.~\ref{fig:Statistics}a) and b) show for each sector $q$ the frequency of $\text{Tr}_{q}$ evaluation. 

Consider first the case of LNO. In contrast to the baseline, it is clear in Fig.~\ref{fig:Statistics}a) that the sector frequencies for the lazy trace evaluation are largely proportional to the sector weights. 
Only a few sectors with $N=7$ to $8$ collect most of the weight, and this not only shows where the large reduction in matrix 
multiplications in Fig.~\ref{fig:Combo}a) comes from, but also why the reduction in floating point operations is even bigger. 
Indeed, the sectors with $N=7$ to $8$ have generally smaller dimension than the ones with $N=4$ to $6$ which are not calculated 
most of time in the lazy trace evaluation.

Given their negligible sector weights, it would also be possible in principle to just drop the sectors with $N=0$ to $3$. However, the gain from this is small since 
these sectors have rather small dimension. Dropping the sectors with $N=4$ to $6$ involves more important approximations so one would need careful checks that the truncated sectors do not affect the results. The lazy trace evaluation avoids the calculation of these sectors 
most of time and there is no approximation involved.

Moving to the case of FeTe in Fig.~\ref{fig:Statistics}b), one notices that the sector weights are more uniformly distributed. There are fewer sectors with extremely small weights. Hence the lazy trace evaluation does not give a substantial speedup. The skip lists on the other hand still reduce the number of matrix multiplications.   

\section{Discussion and Conclusion}

Quantum Monte Carlo algorithms generally involve multiplications of large matrices. In the case of the strong-coupling based CT-HYB algorithm, this is a limiting factor. When updates generate new configurations that have a large probability of being rejected, we have shown that an efficient way of speeding up the algorithm is to first choose the random number and then use matrix norms to bound the Metropolis rejection/acceptation probability. This is called lazy trace evaluation. Skip lists on the other-hand provide a way to store intermediate matrix products and avoid in all circumstances the recomputation of some of the matrix-products. The combination of both algorithms, lazy skip lists, provides a robust algorithm that guarantees large speedups when the trace evaluation takes a large fraction of the computing time. 

The speedup of the trace evaluation achieved with the lazy skip lists algorithm is such that parts of CT-HYB that usually take negligible time compared with the evaluation of the trace, for example measurements, calculation of determinants etc., can now become the limiting factor. 


The tree structure introduced in Ref.~\onlinecite{GullThesis:2008, CTQMCRev} transforms an $O(k)$ to an $O(\log k)$ problem where $k$ is the order in perturbation theory. This substantial gain in speed also applies to skip list. The multiplication algorithm where multiple insertions are done before products are recomputed, as presented in Sec.~\ref{sec:SkipListB} and Sec.~\ref{sec:LazySkipListA}, could be implemented in binary search trees\cite{GullThesis:2008, CTQMCRev} as well. We find skip lists however easier to implement for at least two reasons: first they use simple probabilistic rebalancing rather than explicit rebalancing by tree rotations; second, a linked list is more natural for a product of operators and propagators than a binary tree. 
For the same reasons, skip lists facilitate the exploration of new updates such as exchanging sub-sequences of operators. Also, skip lists allow control of memory requirements by changing the probability $p$ to add a level to an inserted bar after an update. We have not discussed further improvements in speed that can be obtained by using the associative property of matrix multiplication to speedup the calculation of products of rectangular matrices, or many other possible optimizations that are dependent on computer architecture, such as caches, parallelism etc.

It has also been proposed to use Krylov space methods to calculate the trace.~\cite{WernerLauchli:2009} For large enough systems, this approach should be the most efficient one, but for cases of practical interest it might not be. There are optimizations for both the Krylov and the matrix formulation. We first compare the two formulations without optimizations. To this end we consider the number of operations involved in applying both a creation/annihilation operator and a propagator to a state represented by a vector of dimension $d$ in a given symmetry sector. In the matrix formulation, the expensive operation comes from the creation/annihilation operators and costs $d^2$ operations. In the Krylov formulation, the expensive operation is the application of a propagator to a state: it costs the number of operations $N_H$ involved in the application of the Hamiltonian to a state, times the number of Krylov steps $m$. This scales like like $m \times d$. Indeed, taking the product of $d$ and the number of terms in the second quantized Hamiltonian $n_H$ is one way of estimating $N_H$. Another estimate, which is necessarily smaller, is obtained by actually counting the number of non-vanishing elements in the Hamiltonian matrix (of order $d$). Proceeding here with this last estimate for an $f$-shell system in a tetragonal environment, the sector with the biggest dimension has $d=313$ and $N_H=17077$ . The relevant ratio to compare the two approaches in this specific case is thus $m\times 17077/(313)^2 = m \times 0.174$.  It was found in Ref.~\onlinecite{WernerLauchli:2009} that $m$ can be small. However, highly optimized libraries are available when memory is accessed in a regular way, as in the matrix formulation, while the memory access is irregular in the Krylov algorithm. Hence we think that the matrix formulation without the optimizations discussed in this paper can be as fast as the Krylov formulation, even for typical f-shell impurities. Practical implementations must be compared to decide.

For the general case, note that while the lazy trace idea can be applied to the Krylov algorithm, it is less clear that one can implement skip list for this algorithm. Hence, while the Krylov algorithm needs to be repeated $k$ times for an order $k$ term in perturbation theory, the skip-list (or binary tree~\cite{GullThesis:2008,CTQMCRev}) algorithm allows us to change that factor to $\log(k)$. Other optimizations of the Krylov algorithm have been proposed recently~\cite{WernerDolfiOptimization:2014}.

Some of the ideas developed here can be directly applied to other problems treated by Monte Carlo methods. For example the rejection method based on bounds (see Fig.~\ref{fig:bounding}) can be applied to classical Monte-Carlo simulations for spins with long-range interactions:~\footnote{This was suggested by Hiroaki Ishizuka as a further application of our rejection method based on bounds.} Take an Ising spin system and consider a single spin-flip Monte Carlo update. The energy associated with this spin can be bounded by  
\begin{equation}
  E_{i,[\min ,\max ]}=S_{i}\sum\limits_{j\le R}J_{i,j}S_{j}\pm
S_{i}\sum\limits_{j>R}\left\vert J_{i,j}\right\vert .
 \end{equation}
The bounds can be refined by successively increasing the range $R$. The sums over absolute values of exchange constants need to be calculated only once. Similar problems are encountered in spin-ice models with dipolar interactions,~\cite{MelkoGingrasReview:2004} ordered and/or random spins with both dipolar and RKKY interactions. Other schemes relying on different ideas also exist and may be faster.~\cite{LuijtenLongRangeInteractions:1995} But this remains to be tested.

 Speedups by factors in the hundreds that can be achieved with the lazy skip lists algorithm will bring new physical regimes in correlated electronic-structure calculations and cluster generalizations of dynamical mean-field theories within reach of computational power. Applications of such methods extend as far as molecular biology.~\cite{WeberHemoglobin:2013}

\acknowledgments
We are grateful to M. Boninsegni, A. Del Maestro,  M. Gingras, E. Gull, H. Ishizuka, R. Melko, O. Parcollet, M. Troyer, P. Werner, and especially to S. Allen, D. S\'en\'echal, and J. Goulet  for useful discussions. This work has been supported, by the Natural Sciences and Engineering Research Council of Canada (NSERC), by MRL grant DMR-11-21053 (C.Y.) and KITP grant PHY-11-25915 (C.Y.), by NSF grant DMR-0746395 (K.H.) and by the Tier I Canada Research Chair Program (A.-M.S.T.). The ALPS libraries~\cite{Albuquerque:2007, TroyerALPS_ComputerScience:1998} were used in the code. Simulations were run on computers provided by CFI, MELS, Calcul Qu\'ebec and Compute Canada.

\appendix

\section{Trace Bounds via Matrix Norms}
\label{apdx:trace}

Different matrix norms give different bounds for the magnitude of the trace of a
matrix product. We consider here induced norms  
\begin{equation*}
\lVert A \rVert_p := \max_{\lVert x \rVert_p = 1} \lVert Ax \rVert_p,
\end{equation*}  
where $A\in \mathbb{R}^{N\times N}$, $x\in \mathbb{R}^N$ and $\lVert x \rVert_p := (\sum_i |x_i|^p)^{1/p}$ with $p\geq 1$,
and the Frobenius norm
\begin{equation*}
\lVert A \rVert_F := \bigg (\sum_{ij} A_{ij}^2\bigg )^{\frac{1}{2}}.
\end{equation*}

\subsection{Induced Norms}
For the induced norms, one obtains  $|A_{ii}| \le \lVert A e_i \Vert_p \le \lVert A \rVert_p$, where $e_i$ is the standard basis of  $\mathbb{R}^N$,  and hence 
\begin{equation*}
|\text{Tr} A | \le N \cdot \lVert A \rVert_p.
\end{equation*}
This immediately generalizes to a product 
\begin{equation}
\label{equ:induced}
\norm{\Tr \prod_{l=1}^n A_l } \le \min\{N_l\} \cdot \prod_{l=1}^n \lVert A_l \rVert_p
\end{equation}
of rectangular matrices $A_l \in \mathbb{R}^{N_l\times M_l}$, since induced norms are sub-multiplicative. From the cyclicity of the trace,
the pre-factor in Eq.~\ref{equ:norm} becomes $C = \min\{N_l\} = \min\{ M_l\}$, the
minimal row or column dimension of all the matrices within the product. 

 For a propagator $P_\tau$, written in the eigenbasis, one obtains $\lVert P_\tau \rVert_p = \exp(-\tau E_0)$, where $E_0$ is the smallest eigenvalue. These norms are hence well suited for the lazy trace evaluation in Sec.~\ref{sec:Lazy}. Especially convenient is the spectral norm ($p = 2$). This norm is one for annihilation or creation operators since
\begin{equation*}
\lVert d \rVert_2 = \max_{\olap{\psi}{\psi} = 1} \sqrt{\me{\psi}{d^\dagger d}{\psi}} = 1
\end{equation*}
by the Pauli principle, and only the exponentials of the propagators enter into the bound given in equation (\ref{equ:induced}).

\subsection{Frobenius Norm}
For the Frobenius norm, Cauchy-Schwarz states
\begin{equation*}
|\text{Tr}A B| \le \lVert A \rVert_F \cdot \lVert B \rVert_F, 
\end{equation*}
and as the Frobenius norm is sub-multiplicative
\begin{equation}
\label{equ:Frobenius}
\norm{\Tr\prod_{l=1}^n A_l} \le \prod_{l=1}^n \mnorm{A_l}_\text{F},
\end{equation}
where $n \ge 2$. The Frobenius norm is numerically cheap, so equation (\ref{equ:Frobenius}) can be used for the lazy skip lists in Sec.~\ref{sec:LazySkipListB}. Other numerically cheap choices are the induced norms with $p=1$ and $p=\infty$. 


%

\end{document}